\documentclass[aps,a4paper,showpacs,showkeys]{article}
\usepackage[twoside,top=2cm,bottom=2cm,left=2cm,right=2cm]{geometry}
\usepackage[utf8]{inputenc}	%lettere accentate da tastiera
\usepackage{authblk} % Afiliazione autori
\usepackage{epsfig}
\usepackage{amsmath}
\usepackage{amsfonts}
\usepackage{amssymb}
\usepackage{graphicx}
\usepackage{colordvi}
\usepackage{widetext}
\usepackage{float}
\usepackage[title]{appendix}
\date{}
\begin{document}
	\author{ Filippo Maimone$^{(1)}$\footnote{%
			e-mail address: \textit{filippo.maimone@gmail.com}}, Adele Naddeo$^{(2)}$\footnote{%
			e-mail address: \textit{anaddeo@na.infn.it}}, Giovanni Scelza$^{(1)}$\footnote{%
			e-mail address: \textit{lucasce73@gmail.com}}}
	
	\affil{\small{\textit{$^{(1)}$ Associazione Culturale ``Velia Polis", via Capo di Mezzo, $84078$ Vallo della Lucania, Italy}}}
	\affil{\small{\textit{$^{(2)}$INFN, Sezione di Napoli, C. U. Monte S. Angelo, Via Cinthia, $80126$ Napoli, Italy}}}

	\title{Interaction between Everett worlds and fundamental decoherence in Non-unitary
		Newtonian Gravity }
	\maketitle
	
	\begin{abstract}
		It is shown that Non-unitary Newtonian Gravity (NNG) model admits a simple
		interpretation in terms of Feynman path integral, in which the sum over all
		possible histories is replaced by a summation over pairs of paths. Correlations between different paths are allowed by a fundamental decoherence mechanism of gravitational origin and can be interpreted as a kind of communication between different branches of the wave function. The
		ensuing formulation could be used in turn as a motivation to introduce
		Non-unitary Gravity itself.
	\end{abstract}
	
	
	\section{Introduction}
	
	Today there is a growing interest on the topic of intrinsic decoherence \cite{Stamp}, as a phenomenon which, on one hand is analogous to the well known environmental decoherence \cite{zurek1,Joos} in destroying phase coherence, but on the other hand is very different because it is not the result of averaging over an environment. As such it is intrinsic to Nature and may arise in principle also in isolated systems. But it is also expected to lead to an entropy growth, as a consequence of the entanglement between the system and something which plays the role of an environment. Of course, this irreversible process is in sharp contrast with the deterministic unitary evolution of Quantum Mechanics (QM) governed by Schroedinger equation, which means that intrinsic decoherence is involved in the so called measurement problem in QM as well as in the related issue of emergence of classicality from the quantum world. 
	
	Many different proposals of modification of QM have been put forward in the last 50 years, ranging from the inclusion of nonlinear terms \cite{weinberg1,weinberg2} or a random noise source \cite{grw,pearle,bassi,csl1,csl2} in Schroedinger equation, to the suggestion of a possible role of gravity in the collapse of the wave function \cite{karol,frenkel,frenkel1,diosi0,diosi,diosi1,penrose0,penrose1}. Some of these models include free parameters as new constants of Nature, while some others have been recently falsified by accurate experiments \cite{gransasso}. In particular the so called Diosi-Penrose gravity induced collapse model, that has no free parameters, predicts a certain amount of energy non conservation via spontaneous emission of X-rays, a feature that has not been observed. The idea of a gravitational decoherence \cite{grDec,gdec2}, in particular, is attractive since gravity is ubiquitous in Nature and gravitational effects depend on the size of objects. As put forward by R. Penrose \cite{penrose0,penrose1}, a conflict emerges when a balanced superposition of two separate wave packets representing two different position of a massive object is considered, signaling a possible inconsistency between the general covariance of general relativity and the quantum mechanical superposition principle. On the other hand the discovery by Hawking that black holes radiate leading to a highly mixed state of the radiation field after the complete evaporation, looks like a strong indication towards non-unitary time evolution of state vectors within a future quantum theory of gravitation \cite{hawk0,hawk1}. This adds further evidence to the intrinsic character of gravitational decoherence as well. 
	
	More recent proposals assume as source of decoherence fluctuations deriving from gravitational waves, gravitons or metric fluctuations, which obey Einstein equations and can be identified with transverse traceless perturbations \cite{gravitons1,gravitons2}. These differ from models of gravitational decoherence in which the perturbations of the metric don't satisfy linearized Einstein equations but are described by a stochastic process \cite{kok,asprea}. An alternative route to decoherence, always related to gravity, makes it to descend from time dilation and doesn't require modifications of quantum mechanics \cite{pik1,pik2}. In fact it has been shown how time dilation is effective in producing a universal coupling between internal degrees of freedom and the center of mass of a composite particle, whose result is decoherence of the particle's position, even in the absence of an external environment. Finally it is worth to mention gravitationally induced decoherence models based on specific quantum gravity models. Here the starting point is the specific classical model which is quantized and then, within the quantum model, a master equation is derived in order to describe system's dynamics in the presence of gravitational decoherence. See for instance the model in Ref. \cite{ashtekar}, inspired by loop quantum gravity and constructed in terms of Ashtekar variables.
	
	On the experimental side huge advances in technology have been carried out in the last decade in order to test the predictions of models of gravitational decoherence. Viable platforms for large-mass experiments include non-interferometric opto- and magneto-mechanical systems \cite{opto,opto1} and matter-wave interferometers with molecules and nanoparticles \cite{opto2}, while ground-to space optical links, such as the entangled-photon pairs of the Space QUEST mission \cite{quest} and the long-baseline quantum links proposed in the context of the DSQL mission \cite{deep}, are currently pursued because of the peculiar experimental conditions with respect to the ground (see for a detailed account Ref. \cite{microgravity} and references therein). Finally, a novel underground experiment in operation at Gran Sasso National Laboratory \cite{gransasso} is very promising as well thanks to a reduced exposition to cosmic radiation.
	
	To address the quest for a viable model of gravitational decoherence that is based on expected general features and doesn't rely on the presence of external gravitational fields or on specific theories of quantum gravity, a consistent proposal was put forward by De Filippo twenty years ago and then further developed \cite{DeFMaimPRD,DeFMaimAIP,SergioAdele}. At variance with other attempted non unitary modifications of QM, De Filippo's one, namely Non-unitary Newtonian Gravity (NNG), is non-Markovian and guarantees energy conservation. It was obtained as the non-relativistic limit of a non-unitary variant of higher derivative gravity, which makes the theory classically stable while providing a mechanism to tame singularities \cite{DeFMaimPRD}. Its non-unitary dynamics results from a unitary one of a meta-system, built of the physical system and a replica of it, these two subsystems interacting with each other only via gravity. The model treats on an equal footing mutual and self-interactions, which, according to some authors, are believed to produce wave function localization and/or reduction \cite{karol,frenkel,frenkel1,diosi0,diosi,diosi1,penrose0}. In particular, for ordinary condensed matter densities, it exhibits a localization threshold at about $10^{11}$ proton masses, above which self-localized center of mass wave functions exist. This means that initially pure states evolve in time into an ensemble of localized states, consistently with the expectation that, even for an isolated system, an entropy growth takes place due to the entanglement between observable and hidden degrees of freedom via gravitational interaction \cite{Filippo1,Filippo2,noi1,noi2}. In particular the energy pumping into the system, necessary to get localization, is only due to quantum fluctuations and to the natural conversion of gravitational potential energy into kinetic energy.
	
	Another interesting feature of De Filippo's model is its field-theoretical formulation, obtained by a Hubbard-Stratonovich transformation of the gravitational interaction \cite{DeFMaimPRD}, which allows to obtain the well known nonlinear Newton-Schr\"{o}dinger equation \cite{penrose1, Bahrami} by gravitationally coupling $N$ exact copies of the physical system and then taking the limit $N \rightarrow \infty$ \cite{SergioNS}. It has been also shown how, for any finite value of $N$, the same model is completely free from the usual causality violation problems that plagued Newton-Schr\"{o}dinger model \cite{naturalcure}; this happens thanks to the intrinsic mechanism of spontaneous state reduction of the model, absent in the Newton-Schr\"{o}dinger limit. Furthermore, since within NNG density matrix plays a fundamental role, amounting to the most complete characterization of a physical system's state, the
	Everett Many World Interpretation appears to be the most natural conceptual framework of that theory. Here, in agreement with some other non linear modifications of QM \cite{Polchinski}, the possibility of constructing an Everett phone connecting different branches of the wave function has been shown, though its basic mechanism appears to be strongly inhibited. In fact the
	theory gets rid of the huge number of branches continuously forming by
	turning each time the macroscopic states superposition into ensembles of
	localized states through gravitational self-interaction.
	
	In this work our aim is to give a novel interpretation of the above mentioned Everett phone by introducing a path integral formulation of De Filippo's model \cite{FeynmanHibbs}. As a peculiar feature, we find that the sum over all possible histories gets replaced by a summation over pairs of paths, \textit{talking together} via correlations, which are induced by a fundamental decoherence mechanism of gravitational origin. Such correlations can be interpreted as a kind of communication between different branches of the wave function in agreement with our previous findings \cite{naturalcure}.
	
	The plan of the present work is as follows. In Section 2 we give a brief survey of the relevant features of De Filippo's model, while Section 3 contains its path integral formulation for a simple system made of a homogeneous spherical particle with radius $R$. As a concrete example the case of microscopic-to-mesoscopic systems is explicitly worked out, showing that in such a range the contribution of non-unitary gravity amounts to a correction with respect to ordinary QM. In Section 4 an evaluation of this correction is carried out in the framework of a COW-like experiment \cite{COW}. Finally some perspectives of this work are outlined.
	
	\section{Nonunitary Newtonian gravity model: a brief survey}
	
	In this Section a brief introduction to the NNG model and its relevant features is given, by resorting to a second quantization formalism that allows for a general formulation \cite{DeFMaimPRD,DeFMaimAIP,SergioAdele}.
	
	The general idea underlying NNG model is to start from a gravity-free system and to make it to interact with a copy of it (which plays the role of a bath) via gravitational interaction. Then, upon tracing out unobservable (i.e. bath) degrees of freedom, a non unitary dynamics is obtained, which includes both the usual aspects of classical gravitational interactions and a kind of fundamental decoherence, which is expected to play a role in the emergence of classicality for macroscopic bodies. In this way the evolution of an initial pure state of the system into a mixture of localized states takes place, accompanied by a fundamental entropy growth \cite{Filippo1,Filippo2,noi1,noi2}.
	
	Here we present the general formulation, in which a system is put in interaction with $N-1$ hidden copies of it. To this end, let $H[\psi ^{\dagger },\psi ]$ denote the second
	quantized non relativistic Hamiltonian of a finite number of particle
	species, like electrons, nuclei, ions, atoms and/or molecules, according to
	the energy scale. For notational simplicity, $\psi ^{\dagger },\,\psi $
	denote the whole set $\psi _{j}^{\dagger }(x),\,\psi _{j}(x)$ of
	creation-annihilation operators, \textit{i.e.} one couple per particle
	species and spin component. This Hamiltonian includes the usual
	electromagnetic interaction accounted for in atomic and molecular physics.
	Gravitational interactions as well as self-interactions are incorporated by
	introducing a color quantum number $\alpha =1,2,\dots ,N$, in such a way that
	each couple $\psi _{j}^{\dagger }(x),\,\psi _{j}(x)$ is replaced by $N$
	couples $\psi _{j,\alpha }^{\dagger }(x),\,\psi _{j,\alpha }(x)$ of
	creation-annihilation operators. 
	
	The resulting overall Hamiltonian, including
	gravitational interactions and acting on the tensor product $\otimes
	_{\alpha }F_{\alpha }$ of the Fock space of the $\psi _{\alpha }$ operators,
	is given by
	\begin{equation}
		H_{G}=\sum_{\alpha =1}^{N}H[\psi ^{\dagger },\psi ]-\frac{G}{N-1}%
		\sum_{j,k}m_{j}m_{k}\sum_{\alpha <\beta }\int dx\,dy\frac{\psi _{j,\alpha
			}^{\dagger }(x),\,\psi _{j,\alpha }(x)\psi _{k,\beta }^{\dagger }(y),\,\psi
			_{k,\beta }(y)}{|x-y|},
	\end{equation}
	
	where here and henceforth Greek indices denote color indices, $\psi _{\alpha
	}\equiv (\psi _{1,\alpha },\psi _{2,\alpha },\dots ,\psi _{N,\alpha })$ and $%
	m_{i}$ denotes the mass of the $i-$th particle species, while $G$ is the
	gravitational constant. While the $\psi _{\alpha }$ operators obey the same
	statistics as the original operators $\psi $, we take advantage of the
	arbitrariness pertaining to distinct operators and, for simplicity, we choose
	them commuting with one another: $\alpha \neq \beta \Rightarrow \lbrack \psi
	_{\alpha },\psi _{\beta }]_{-}=[\psi _{\alpha },\psi _{\beta }^{\dagger
	}]_{-}=0$. 
	
	The meta-particle state space $S$ is identified with the subspace
	of $\otimes _{\alpha }F_{\alpha }$ including the meta-state obtained from the
	vacuum $\Vert 0\rangle\rangle =\otimes _{\alpha }|0\rangle _{\alpha }$
	applying operators built in terms of the product $\prod_{\alpha =1}^{N}\psi
	_{j,\alpha }^{\dagger }(x_{\alpha })$ and symmetrical with respect to
	arbitrary permutations of the color indices, which, as a consequence, for
	each particle species, have the same number of meta-particles of each color.
	This is a consistent definition since the time evolution generated by the
	overall Hamiltonian is a group of (unitary) endomorphism of $S$. If we
	prepare a pure $n-$particle state, represented in the original setting,
	excluding gravitational interactions, by
	\begin{equation*}
		|g\rangle \doteq \int d^{n}x\,g(x_{1},x_{2},\dots ,x_{n})\psi
		_{j_{1}}^{\dagger }(x_{1}),\psi _{j_{2}}^{\dagger }(x_{2})\dots ,\psi
		_{j_{n}}^{\dagger }(x_{n})|0\rangle ,
	\end{equation*}
	its representative in $S$ is given by the meta-state
	\begin{equation*}
		\Vert g^{\otimes N}\rangle \rangle \doteq \prod_{\alpha }\biggl[\int
		d^{n}x\,g(x_{1},x_{2},\dots ,x_{n})\psi _{j_{1},\alpha }^{\dagger
		}(x_{1}),\psi _{j_{2},\alpha }^{\dagger }(x_{2})\dots ,\psi _{j_{n},\alpha
		}^{\dagger }(x_{n})\biggr]\Vert 0\rangle \rangle .
	\end{equation*}
	As for the physical algebra, it is identified with the operator algebra of
	say the $\alpha =1$ meta-world. In view of this, expectation values can be
	evaluated by previously tracing out the unobservable operators, namely with $%
	\alpha >1$, and then taking the average of an operator belonging to the
	physical algebra. Clearly this should not be viewed as an
	\textit{ad hoc} restriction of the observable algebra. Indeed, once the constraint
	restricting $\otimes _{\alpha }F_{\alpha }$ to $S$ is taken into account,
	in order to get an effective gravitational interaction between particles of
	one and the same color, the resulting state space does not contain states
	that can distinguish between operators of different color. Restricting the algebra to that of $\psi _{1}$ operators is the only way to accommodate a faithful representation of the physical algebra within the meta-state space. Furthermore the resulting constrained\ theory is, by construction, a fully consistent QM theory.
	
	The general formulation of NNG above introduced allows one to recover in the limiting case $N \rightarrow \infty$ the Newton-Schr\"{o}dinger model \cite{Bahrami} as a result of a well-defined procedure \cite{SergioNS,SergioAdele}. This limit has been shown to completely suppress quantum fluctuations while preserving mean field features. On the other hand a simple thermodynamic argument points to take $N=2$ \cite{noi2}, thus giving a free-parameter model.

	\section{Path integral formulation of NNG model in a simple case}
	
	Let's consider, within NNG model, a system composed
	(for simplicity) by a single body of mass $m$ and study its dynamics by resorting to a path integral formulation \cite{FeynmanHibbs}. We show that the evolution of the system cannot be expressed \ by a summation over a single
	path history, but instead needs to be written as a summation over \textit{%
		pairs} of histories.
	
	The (meta-)amplitude for the particle to go from a position $\mathbf{x}_{a}$
	at time $t_{a}$ to a position $\mathbf{x}_{b}$ at time $t_{b}$, while its
	hidden counterpart is going from $\mathbf{x}_{a}$ to $\widetilde{\mathbf{x}}%
	_{b}$ at the same times, is given by:
	\begin{equation}
		\langle \langle \mathbf{x}_{b},\widetilde{\mathbf{x}}_{b};t_{b}\ \Vert \ \mathbf{%
			x}_{a},\mathbf{x}_{a};t_{a}\rangle \rangle =\int \mathcal{D}[\mathbf{x}%
		\left( t\right) ]\mathcal{D}[\widetilde{\mathbf{x}}\left( t\right) ]~\exp
		\left\{ i\frac{S_{0}\left[ \mathbf{x}\left( t\right) \right] +S_{0}\left[ 
			\widetilde{\mathbf{x}}\left( t\right) \right] +S_{G}\left[ \left\vert 
			\mathbf{x}\left( t\right) -\widetilde{\mathbf{x}}\left( t\right) \right\vert %
			\right] }{\hbar }\right\} .  \label{Eq1}
	\end{equation}\\
	Here $S_{0}$ is the action for the particle, which includes the kinetic
	energy and an external potential added, in the general case, to the halved
	Newtonian gravitational potential. Instead, $S_{G}$ contains the non-unitary
	gravitational interaction and takes the form
	\begin{equation*}
		S_{G}=-\ \int_{t_{a}}^{t_{b}}dt\ V_{G}\left( \left\vert \mathbf{x}\left(
		t\right) -\widetilde{\mathbf{x}}\left( t\right) \right\vert \right) ,
	\end{equation*}
	where $V_{G}$ is the halved gravitational potential between the body and its hidden counterpart, which for a homogeneous
	spherical particle of mass $m$ and radius $R$ is \cite{Filippo2}:
	\begin{equation*}
		V_{G}\left( r\right) =\frac{1}{2}Gm^{2}\left\{ \frac{\theta \left(
			2R-r\right) \ \left( 80R^{3}r^{2}-30R^{2}r^{3}+r^{5}-192R^{5}\right) }{160\
			R^{6}}-\frac{\theta \left( r-2R\right) \ }{r}\right\} ,
	\end{equation*}
	and $\theta$ denotes the Heaviside function. 
	
	Starting from Eq. (\ref{Eq1}), the density matrix $\rho _{P}$ of the particle can be calculated integrating over the space coordinates $\widetilde{{\xi }}$:
	\begin{equation}
		\rho _{P}=\int d^{3}\widetilde{{\xi }}\ \langle \widetilde{\mathbf{%
				\xi }}\left\vert \left\vert \mathbf{\Psi };t_{b}\right\rangle \right\rangle
		\left\langle \left\langle \mathbf{\Psi };t_{b}\right\vert \right\vert 
		\widetilde{\mathbf{\xi }}\rangle ,  \label{roP}
	\end{equation}
	where
	\begin{equation*}
		\left\vert \left\vert \mathbf{\Psi };t_{b}\right\rangle \right\rangle =\int
		d^{3}\mathbf{x}_{b}\int d^{3}\widetilde{\mathbf{x}}_{b}\left\vert
		\left\vert \mathbf{x}_{b},\widetilde{\mathbf{x}}_{b};t\right\rangle
		\right\rangle \ \langle \langle \mathbf{x}_{b},\widetilde{\mathbf{x}}_{b};t\
		\Vert \ \mathbf{x}_{a},\mathbf{x}_{a};t_{a}\rangle \rangle .
	\end{equation*}
	It follows immediately that, because of the coupling term $S_{G}$, the
	expression of $\rho _{P}$, containing all the information that can be
	experimentally accessible, cannot be reduced to a path integral over a
	single history. Indeed inter-path correlations between different branches of the propagator arise, so that the whole transition amplitude is obtained by summing both over the different paths and over correlations between them, or in other words by summing over pairs of paths. Here such correlations are induced by a fundamental decoherence mechanism of gravitational origin, which is built in the NNG model, and can be interpreted as a kind of communication between different branches of the wave function.\footnote{In a different view, instead of a communication between Everett branches, the whole mechanism can be interpreted, in general, in terms of interactions between the ``environments" formed by the bodies' copies and the bodies themselves.} These conclusions are consistent with previous findings, pointing out the appearance of communications among Everett branches of the wave function as a consequence of the introduction of nonlinearities in some modified theories of quantum mechanics \cite{Polchinski,naturalcure}. Furthermore our results are in agreement with existing treatments of intrinsic decoherence mechanisms\cite{Stamp,Stamp1}. Generalization to the case with $n$ particles is straightforward. 
	
	It's important to note that, even if it has been shown that Non-unitary
	Newtonian Gravity model has neat observational signatures (able, in
	principle, to distinguish it from the other proposed models), its
	observational implications are far from being obvious \cite{mirror}. In
	fact they concern mainly measurements of coherences in the basis of
	positions, rather than some probability of occurrence. As an example, let's
	write down explicitly the probability to find the particle at point $\mathbf{%
		X}$ at time $t_{b}\ $:
	\begin{equation*}
		\Pr \left( \mathbf{x}_{b}\equiv \mathbf{X}\text{ at time }t_{b}\right)
		\equiv \rho _{P}\left( \mathbf{X,X;}t_{b}\right) =\int d^{3}\widetilde{%
			\mathbf{\xi }}\ \left\vert \langle \langle \mathbf{X},\widetilde{\mathbf{\xi 
		}};t_{b}\ \Vert \ \mathbf{x}_{a},\mathbf{x}_{a};t_{a}\rangle \rangle
		\right\vert ^{2}.
	\end{equation*}
	For microscopic systems gravity is irrelevant and the term $S_{G}$ can be
	neglected. Then the meta-amplitude factorizes, and due to the normalization
	\begin{equation*}
		\int d^{3}\widetilde{\mathbf{\xi }}\ \left\vert \langle \widetilde{\mathbf{%
				\xi }};t_{b}\mid \mathbf{x}_{a};t_{a}\rangle \right\vert ^{2}=\int d^{3}%
		\widetilde{\mathbf{\xi }}\left\vert \int \mathcal{D}[\widetilde{\mathbf{x}}%
		\left( t\right) ]~\exp \left\{ i\frac{S_{0}\left[ \widetilde{\mathbf{x}}%
			\left( t\right) \right] }{\hbar }\right\} \right\vert ^{2}=1,
	\end{equation*}
	the probability above reduces obviously to a single history expression,
	according to ordinary QM.
	
	Instead, for microscopic-to-mesoscopic systems typically $S_{0}$ is
	comparable or greater than $\hbar $, but we have still $S_{G}\ll \hbar $. Then it is
	possible to calculate the contribution of non-unitary gravity as a small
	correction with respect to ordinary QM. To be specific, we can write the
	probability above as
	\begin{eqnarray}
		&& \Pr \left( \mathbf{x}_{b}\equiv \mathbf{X}\text{ at
			time }t_{b}\right)   \notag \\
		&=&\int d^{3}\widetilde{\mathbf{\xi }}\ \left\vert \underset{A}{\underbrace{%
				\int \mathcal{D}[\mathbf{x}\left( t\right) ]\mathcal{D}[\widetilde{\mathbf{x}%
				}\left( t\right) ]~e^{i\frac{S_{0}\left[ \mathbf{x}\left( t\right) \right]
						+S_{0}\left[ \widetilde{\mathbf{x}}\left( t\right) \right] }{\hbar }}}}+%
		\underset{a}{\underbrace{\frac{i}{\hbar }\int \mathcal{D}[\mathbf{x}\left(
				t\right) ]\mathcal{D}[\widetilde{\mathbf{x}}\left( t\right) ]~S_{G}\left[
				\left\vert \mathbf{x}\left( t\right) -\widetilde{\mathbf{x}}\left( t\right)
				\right\vert \right] \ e^{i\frac{S_{0}\left[ \mathbf{x}\left( t\right) \right]
						+S_{0}\left[ \widetilde{\mathbf{x}}\left( t\right) \right] }{\hbar }}}}%
		\right\vert ^{2}  \notag \\
		&=&\int d^{3}\widetilde{\mathbf{\xi }}\ \left\vert A+a\right\vert ^{2}\simeq
		\int d^{3}\widetilde{\mathbf{\xi }}\ \left[ \left\vert A\right\vert ^{2}+2%
		Re\left( Aa^{\ast }\right) \right] ,  \label{Aa}
	\end{eqnarray}
	the last term in square parenthesis representing the expected correction.

	\section{Analysis of a COW-like experiment}
	
	In this Section we deal with the explicit calculation of the correction term provided in Eq. (\ref{Aa}) in a specific context. The aim is to estimate the contribution of NNG to the quantum dynamics.
	
\begin{figure}[H]
		\centering
		\includegraphics[width=0.4\textwidth]{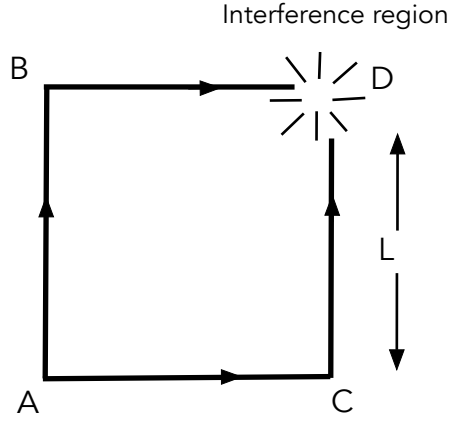}
		\caption{Scheme of the matter interferometry apparatus. A superposition of two wave packets is sent 
			in point A towards B and C respectively, where they are reflected, and then recombine in D.}
		\label{fig1}
	\end{figure}

	To this end let's consider an experiment similar to the famous 1975 experiment by
	Colella, Overhauser and Werner \cite{COW}, using a neutron interferometer
	to detect the Earth's gravitational field effect, as depicted in Figure \ref{fig1}.
	A nearly mono-energetic beam of particles with velocity $v$, that may be
	thermal neutrons or almost spherical macromolecules like fullerene $C_{60}$,
	is split into two parts at a point A. The resulting beams travel towards the mirrors B and C where they are reflected and then are brought together in D at time $T=\frac{2L}{v}$, as
	shown in Fig. 1. Because the size of the wave packets can be assumed to
	be much smaller than the macroscopic dimension of the square loop (that in
	a neutron interferometer is of the order of $10 \ cm$), it is possible to
	consider the trajectories as classic. Let's make the assumption that the two alternative
	paths ABD and ACD are subject to different external potentials (for example
	the Earth's gravitational potential, or an electric potential). Under these conditions the term $%
	\left\vert A\right\vert ^{2}$ in (\ref{Aa}), corresponding to ordinary QM
	amplitudes' summation,  is given by
	\begin{equation*}
		\left\vert A\right\vert ^{2}=\left\vert \frac{e^{i\frac{S_{0}\left[ \mathbf{x%
					}_{ABD}\left( t\right) \right] }{\hbar }}+e^{i\frac{S_{0}\left[ \mathbf{x}%
					_{ACD}\left( t\right) \right] }{\hbar }}}{2}\right\vert ^{2}=\cos ^{2}\left( 
		\frac{\delta }{2\hbar }\right) ,
	\end{equation*}
	where $\delta $ is the overall phase difference in D induced by the
	paths' different potentials. Instead the correction term $Aa^{\ast }$,
	corresponding to the pairs of histories' summation, is given \textbf{by developing explicitly the double path summation of Eq. (\ref{Aa}) on the interferometer arms as}
	\begin{equation}
		Aa^{\ast }=\left( \frac{-iS_{G}^{\left( 0\right) }}{4\hbar }\right) \left\{ 
		\frac{1}{2}+\cos \left( \frac{\delta }{\hbar }\right) +\frac{1}{2}\cos
		\left( \frac{2\delta }{\hbar }\right) +\left( \frac{S_{G}^{\left( 1\right) }%
		}{S_{G}^{\left( 0\right) }}\right) \left[ 1+\cos \left( \frac{\delta }{\hbar 
		}\right) \right] \right\} ,  \label{Aa2}
	\end{equation}
	where
	\begin{equation*}
		S_{G}^{\left( 0\right) }=-T\ V_{G}\left( 0\right) \ \text{\ and \ }%
		S_{G}^{\left( 1\right) }=-2\int_{0}^{T/2}dt\ V_{G}\left( \left\vert \sqrt{2}%
		v\ t\right\vert \right) .
	\end{equation*}
	Since the expression in Eq. (\ref{Aa2}) is purely imaginary, no contribution at all
	comes from the correction term! This happens in spite of the asymmetry between the two
	different paths brought by the external potential.
	
	Then, the lesson to be drawn from the above analysis is that it is almost
	impossible to distinguish between ordinary QM and its pairs of histories
	modification due to NNG, as far as micro/mesoscopic systems and not-very tailored
	experiments are considered. An example of a possible experimental test of NNG, involving mesoscopic masses and currently still beyond our technological reach, can be found in Ref. \cite{mirror}.

	\section{Concluding remarks}
	
	In this work we give a simple path integral formulation of NNG model in which the sum over all possible histories gets replaced by a sum over pairs of paths, thanks to the occurrence of interpath correlations induced by a fundamental decoherence mechanism built in the model.
	
	Our results add further motivation to introduce
	Non-unitary Newtonian Gravity itself. In fact, in the search for a viable low-energy theory
	incorporating in a consistent way (quantum) self-gravitational effects,
	several indications point to fundamental decoherence as a general
	characteristic of the theory. Then, according to the analysis performed in
	Ref. \cite{Stamp}, a natural way to guess the form of such a modification of
	QM is to write the Feynman propagator by a sum over pairs of histories
	rather than over single histories. This is inspired by the theory of open
	quantum systems (i.e., of environmental decoherence) in which the central quantity to compute is the Feynman-Vernon influence functional \cite{FeynmanHibbs}, amounting to a double path integral. In our case the
	fundamental view of an interaction between the different branches of the
	particle itself is considered. On the other hand, the interaction between
	Everett branches has to be taken into account when dealing with the
	measurement problem, since a complete independence of the Everett worlds
	cannot account for a whatsoever mechanism of objective reduction of the
	quantum state by means of a (macroscopic) measuring apparatus (see Ref.
	\cite{naturalcure} for a more detailed discussion on this point).
	
	The ensuing picture of a \textit{soft} version of the Everett Many World Theory avoids the continuous macroscopic splitting of ourselves, while leaving the room for a real quantum parallelism in the mesoscopic domain. 
	
	%, including some structures of our brains. The growing empirical evidences in the new field of Quantum Cognition, straddling between physics and psychology \cite{pho}, seem indeed to point to this last possibility.
	
	A further remark is in order concerning the possible experimental detection of fundamental decoherence signatures within NNG model. As shown by our analysis of a COW-like experiment, the choice of the system as well as the design of the apparatus are very challenging. Nevertheless, a successful experimental protocol could also provide the means to discriminate intrinsic decoherence against ordinary environmental decoherence. Huge technological efforts in this direction, and in general aimed to push further the limits of experimental accessibility towards meso- to macroscopic domain, are currently under way.

	\section*{Acknowledgements}
	
	The authors thank the anonymous referees for critical comments which allowed them to improve the manuscript.

	%\begin{thebibliography}{99}

\end{document}